\documentclass[letterpaper]{article} %
\usepackage{arxiv}  %
\usepackage{times}  %
\usepackage{helvet}  %
\usepackage{courier}  %
\usepackage[hyphens]{url}  %
\usepackage{graphicx} %
\usepackage{natbib}  %
\usepackage{caption} %
\usepackage{relsize}
\usepackage{algorithm}
\usepackage{algpseudocode}

\algrenewcommand\algorithmicrequire{\textbf{Input:}}
\algrenewcommand\algorithmicensure{\textbf{Output:}}

\usepackage{newfloat}
\usepackage{listings}
\DeclareCaptionStyle{ruled}{labelfont=normalfont,labelsep=colon,strut=off} %
\floatstyle{ruled}
\newfloat{listing}{tb}{lst}{}
\floatname{listing}{Listing}
\title{Confounding-Robust Deferral Policy Learning}
\author{
    Ruijiang Gao\textsuperscript{\rm 1}, Mingzhang Yin\textsuperscript{\rm 2}
}
\affiliations{
    \textsuperscript{\rm 1}Naveen Jindal School of Management, University of Texas at Dallas, Richardson, TX 75082\\
    \textsuperscript{\rm 2}Warrington College of Business, University of Florida, Gainesville, FL 32611\\
    ruijiang.gao@utdallas.edu, mingzhang.yin@warrington.ufl.edu
}

\usepackage{bibentry}

\usepackage{tikz}
\usetikzlibrary{positioning}
\usepackage{pgflibraryshapes}

\usepackage{graphicx}
\usepackage{color}
\usepackage{enumitem}

\usepackage[utf8]{inputenc} %
\usepackage[T1]{fontenc}    %

\usepackage{url}            %
\usepackage{booktabs}       %
\usepackage{amsfonts}       %
\usepackage{nicefrac}       %
\usepackage{microtype}      %
\usepackage{xcolor}         %

\usepackage{amsmath,amsthm}
\usepackage{bbold}

\usepackage{cleveref}
\Crefname{equation}{Eq.}{Eqs.}
\Crefname{figure}{Fig.}{Figs.}
\Crefname{tabular}{Tab.}{Tabs.}

\newtheorem{theorem}{Theorem}
\newtheorem{corollary}{Corollary}[theorem]

\newtheorem{assumption}{Assumption}

\newcommand{\ie}{\textit{i}.\textit{e}., }
\newcommand{\eg}{\textit{e}.\textit{g}., }

\usepackage{graphicx}
\usepackage{subcaption}

\begin{document}

\maketitle

\begin{abstract}
Human-AI collaboration has the potential to transform various domains by leveraging the complementary strengths of human experts and Artificial Intelligence (AI) systems. 
However, unobserved confounding can undermine the effectiveness of this collaboration, leading to biased and unreliable outcomes. 
In this paper, we propose a novel solution to address unobserved confounding in human-AI collaboration by employing sensitivity analysis from causal inference. Our approach combines domain expertise with AI-driven statistical modeling to account for potentially hidden confounders. 
We present a deferral collaboration framework for incorporating the sensitivity model into offline policy learning, enabling the system to control for the influence of unobserved confounding factors. 
In addition, we propose a personalized deferral collaboration system to leverage the diverse expertise of different human decision-makers. 
By adjusting for potential biases, our proposed solution enhances the robustness and reliability of collaborative outcomes. 
The empirical and theoretical analyses demonstrate the efficacy of our approach in mitigating unobserved confounding and improving the overall performance of human-AI collaborations. Code and appendix are available at \url{https://github.com/ruijiang81/ConfoundL2D}. 
\end{abstract}

\section{Introduction}
In recent years, policy learning has emerged as a powerful tool for learning and optimizing decision-making policies across a diverse range of applications, including healthcare, finance, and marketing \citep{imbens2024causal}. 
One of the most promising avenues for leveraging machine learning is policy learning on observational data \citep{athey2021policy}, which aims to infer optimal decision rules from historical data without the need for costly randomized experiments. Observational data, generated from real-world systems, is abundant and easily accessible, making it an attractive source for training models that can guide policy decisions. 

Many algorithms have been proposed for efficient policy learning from observational data \cite{joachims2018deep,gao2021human,kallus2021more}, usually under the  \textit{unconfoundness} assumption. It assumes %
no hidden confounders that simultaneously influence both the treatment assignment and individual outcomes \citep{rubin1974estimating}. This assumption is  defensible in certain domains such as automated recommendation or pricing systems \cite{biggs2021loss} where we have full control of the historical algorithm, but may rarely hold true for domains where the observational data is generated by human decision-makers. 

Consider a healthcare scenario, where observational data is generated by human experts as the electronic health records (EHRs). These records contain a wealth of information about patients' medical histories, treatments, and outcomes to inform policy learning for personalized medical interventions. However, human experts, such as physicians, may seek additional information when making decisions about patient care, such as the patient's lifestyle, mental well-being, or other contextual factors like the bedside information that might influence their decision-making process as well as the patient's health outcomes. This additional information, though crucial for decision-making, may not be recorded in the EHRs, leading to potential confounding issues in the observed data. For example, a physician may prescribe specific medication to patients with a specific lifestyle, so the observed treatment might be confounded by the potentially unrecorded lifestyle factor. 
In this case, the unobserved confounding can result in suboptimal actions and reduce the reliability of learned policies. In the causal inference literature, the marginal sensitivity model has been proposed under the unmeasured confounding to bound the possible value of the true propensity score \citep{tan2006distributional}. This idea was recently applied to policy learning without humans in the loop \citep{kallus2021minimax}.

In this paper, we propose a human-AI collaboration system 
that learns a policy robust to unmeasured confounding. The system uses a deferral component to decide task allocations to human experts or algorithms. 
The learned policy improves over the algorithm-alone and the human-alone approaches. 
Supposing the historical data in our motivating example are all generated by human decision-makers, an AI-only algorithm is likely to be inferior to humans in cases where external information, such as patients' lifestyles, 
is necessary for optimal decision-making. The benefit of human involvement stands out in the confounding setting as human decision-makers are adept at making choices based on (unobserved) confounding factors \cite{holstein2023toward}. In contrast, a human-only system often incurs a high operation cost. An essential problem is how to jointly learn a rule to choose decision-makers and a rule to assign treatment once the AI system is chosen, especially when the observed data have missing confounders. We refer to this problem as \textit{deferral collaboration under unobserved confounding} \cite{gao2023learning}. ~\looseness=-1

By adopting a human-AI collaborative approach, we can alleviate the impact of these unobserved confounders in the traditional deferral collaboration.
In addition, the external information of human experts can be leveraged by the AI system beyond the observed data to obtain a more accurate estimate of the optimal policy. 
This collaborative framework ensures that the learned policies better account for the missing confounders and yield more reliable decision-making. 

We make the following contributions in the paper: 
We are the first to propose leveraging the learning-to-defer framework to tackle the policy learning under unobserved confounding problem.
We propose a novel algorithm for the problem of \textit{deferral collaboration under unobserved confounding}, where our algorithm works under an uncertainty set over the nominal propensity scores.   
The proposed algorithm leverages human decision-makers who have the capacity to acquire additional unrecorded information to aid their decision-making and a trained algorithmic policy. Theoretically, we prove it is guaranteed to offer policy improvements over a baseline policy based only on the available features or only on the incumbent human policy.
In addition, we generalize our algorithm to personalized settings where each instance can be routed to a specific human decision-maker by exploiting the diverse expertise of humans. 
We theoretically and empirically validate the efficacy of the proposed method.

\section{Related Work}

 We consider human-AI collaboration as a decision-making problem in contrast to the predictive problem mostly considered by the extant human-AI systems. Related to policy learning on the observational data, we relax the untestable unconfoundedness assumption using sensitivity analysis from causal inference.  The proposed human-AI algorithm is related to several threads of literature.
 
\vspace{1mm}
\noindent \textbf{Policy Learning with Unconfoundness.}
Deducing an optimal personalized policy from offline data has been extensively explored in various domains, including e-commerce, contextual pricing, and medicine 
\citep{dudik2014doubly,athey2017efficient,kallus2018balanced,kallus2019classifying,gao2021enhancing,sondhi2020balanced,swaminathan2015counterfactual}. These studies usually assume the historical data were generated by a previous decision-maker, focusing on estimating treatment effects or optimizing an algorithmic policy without human involvement. %
It is yet underdeveloped for scenarios that could benefit from a combined human-AI team to enhance decision performance. ~\looseness=-1

\vspace{1mm}
\noindent \textbf{Sensitivity Analysis.}  Sensitivity analysis is widely used in causal inference that evaluates unconfoundedness assumption \cite{cornfield1959smoking}. 
A popular framework models the confounding effect on the treatment assignment nonparametrically. 
Among them, the marginal sensitivity model (MSM), generalizing the Rosenbaum sensitivity model \citep{rosenbaumn2002observational}, assumes a bound on the odds ratio of the propensity score conditional on the observed variables and a true propensity score conditional on all the confounding variables \citep{tan2006distributional}. 
The MSM has been applied in estimating heterogeneous treatment effects  \citep{yin2021conformal,Jin2022}, robust optimization \citep{Namkoong2022Minimax,guo2022partial}, and policy learning without human in the loop \citep{kallus2021minimax}. 
We adopt the MSM to quantify the deviation of unconfoundedness %
in the context of human-AI collaboration. ~\looseness=-1

\vspace{1mm}
\noindent \textbf{Human-AI Collaboration.}
Recent studies on human-AI collaboration methods improved classification performance, such as accuracy and fairness, by capitalizing on the complementary strengths of humans and AI \citep{bansal2019case,ibrahim2021eliciting,wolzcynski2022learningadvise}. We focus  on the setting without human-AI interaction, where decisions are made by either a human or an algorithm.
Previous research has also addressed the task of routing instances to either a human or an algorithm \citep{madras2018predict,wilder2020learning,raghu2019algorithmic,de2020regression,wang2020augmented}. 
The primary distinction between these studies and ours is that they explore contexts where the AI's learning task is a conventional supervised classification task while we focus on policy learning. 
\citet{gao2021human,gao2023learning} study how to design a deferral collaboration system similar to ours under the unconfoundness assumption, and does not consider the bias due to unmeasured confounding that is often leveraged by humans \cite{holstein2023toward}.  ~\looseness=-1

\section{Confounding-Robust Deferral Policy}

\subsection{Problem Setup}
Assume we have access to the observed  tuples $\{X_i, T_i, Y_i\}_{i=1}^N$, 
where the covariates $X_i\in\mathcal{X}$, the treatment arm $T_i \in \{0,\cdots, m-1\}$, and a scalar outcome $Y_i\in\mathbb{R}$. Using the potential outcome framework, we assume $Y_i = Y_i(T_i)$, \ie %
the SUTVA assumption \cite{rubin1980randomization}. We consider $Y$ as the risk and aim to minimize the risk aggregated over the population. 
In practice, humans often utilize additional information for decisions. For example, a customer service representative may use emotion information in the phone call to decide the compensation plan, but such information cannot be recorded in the past due to the legacy computer system.  We assume such unobserved confounder is $U_i$ 
and the unconfoundedness assumption would hold if we account for both $U_i$ and $X_i$. The $U_i$ can be postulated as the unmeasured covariate or as the unobserved potential outcome itself, \ie $U_i = Y_i(t)$ \cite{zhao2019sensitivity}.
The data is generated by the human decision maker with the \emph{behavior policy} $\pi_0$ as $T_i \sim \text{Categorical}(\pi_0(T_i | X_i,Y_i))$.

Due to the unobserved confounding, the true propensity %
$\pi_0(t|x,y) := P(T=t|X=x,Y(t)=y)$ generally cannot be identified using the observational data alone. 
We can only estimate the nominal propensity, denoted as $\tilde{\pi}_0(t|x) := P(T=t|X=x)$. The nominal propensity can be estimated from the observational data using a machine learning classifier such as logistic regression. 
To quantify the difference between the nominal and true propensity scores incurred by confounding, we adopt the MSM \cite{tan2006distributional} to assume an uncertainty set. %

\begin{assumption}[Marginal Sensitivity Model]
\begin{align}
\label{eqn:sensitivity}
    \Gamma^{-1} \leq  \frac{(1-\tilde{\pi}_0(T|X)\pi_0(T|X,Y)}{\tilde{\pi}_0(T|X)(1-\pi_0(T|X,Y))} \leq \Gamma.  
\end{align}
\end{assumption}
The MSM quantifies the deviation from the true propensity scores by the scalar parameter $\Gamma \geq 1$.  When $\Gamma = 1$, it corresponds to the unconfoundness setup. $\Gamma$ can be determined using domain knowledge or estimated using empirical data, which we will discuss in \Cref{sec:improvement}. 

Deferral collaboration \cite{madras2018predict,gao2021human} considers how to evaluate and learn a routing algorithm $\phi : \mathcal{X} \rightarrow [0,1]$  that assigns tasks to 
the human decision-makers or AI system, and 
an algorithmic policy 
$\pi : \mathcal{X} \rightarrow \Delta^m$ that decides the treatment distribution. The element in simplex $\Delta^m$ is the probability over the treatment arms and $\phi(X)$ denotes the probability of routing to humans. 

The routing algorithm is designed to \emph{complement} human decision-makers. A successful deferral collaboration  routes  different instances to the entity that is likely to yield the best reward by $\phi(X)$, and it leverages the policy $\pi(X)$ for the instances routed to the AI. 
The human decision-maker may incur a cost of $C(X)$ for producing a decision on an instance. %

In this paper, we consider a general setting with multiple human decision-makers $H \in \{1,\cdots,K\}$. Accordingly, the data is generated by first assigning an instance with covariates $X_i$ to different human decision-makers by the rule $d_0(H_i|X_i): \mathcal{X} \to \Delta^K$. Each human decision-maker $H_i$ chooses the treatment by the behavior policy $\hat{\pi}_{0}(T_i|X_i,H_i, Y_i)$. The observed data become $\{X_i, H_i, T_i, Y_i\}_{i=1}^N$.
The routing algorithm $\phi$ is generalized to $\phi: \mathcal{X}\rightarrow \Delta^{K+1}$ where $\phi(A|X), \phi(H|X)$ means the probability of routing instances to the algorithm and a specific human expert $H$. The goal is to learn an optimal routing algorithm $\phi$ and policy $\pi$ that minimizes the  risk.  The process is illustrated in \Cref{fig:illus2}.

\subsection{Our Method: Deferral Collaboration with Unobserved Confounding}

We first consider the situation of homogeneous human experts who have similar decision performance. The expected team performance can be calculated by the self-normalized Hájek estimator \citep{swaminathan2015self} ~\looseness=-1 
\begin{align}
     \theta&(\pi, \phi)= \mathbb{E}\phi(X)(Y+C(X)) +       \label{eq:first} \\
     & \sum_{t=0}^{m-1}{\mathbb{E}  \frac{\mathbb{I}(T=t)}{\pi_0(T|X,Y)} \pi(T|X) Y (1-\phi(X))} /{\mathbb{E}\frac{\mathbb{I}(T=t)}{\pi_0(T|X,Y)}}. \nonumber
\end{align}
Throughout the paper, without further specification, the expectation is with respect to the underlying data distribution. The first term of \Cref{eq:first} is the cost of assigning to human by $\phi$ and the second term is the cost of assigning to the algorithm with policy $\pi$.  Note that now the propensity score depends on both $X$ and $Y$ because of the unobserved confounding. The equality is because $\mathbb{E}\frac{\mathbb{1}(T=t)}{\pi_0(T|X,Y)}=1$ for every $t$.

Practically, we are often interested in human-AI systems that can outperform either the human, or a candidate algorithmic policy. 
Suppose in addition there is a baseline policy $\pi_c(T|X)$, such as the never-treat policy $\pi_c(0|x)=1$ or a candidate algorithmic policy learned from data, that the proposed human-AI system aims to improve upon.
The objective can be written as the improvement over $\pi_c(T|X)$,
\begin{align}\label{eqn:obj_r}
 & R(\pi, \phi, \pi_c) = \mathbb{E}\phi(X)(Y+C(X)) + \nonumber\\
 &\sum_{t=0}^{m-1}\frac{\mathbb{E}  \frac{\mathbb{I}(T=t)}{\pi_0(T|X,Y)} Y [(1-\phi(X))\pi(T|X)-\pi_c(T|X)]}{\mathbb{E}\frac{\mathbb{I}(T=t)}{\pi_0(T|X,Y)}}.
\end{align}

\tikzset{node/.style={rectangle,fill=gray!10,draw,minimum size=0.8cm,inner sep=0.05cm} }
\tikzset{node/.style={rectangle,fill=gray!10,draw,minimum size=0.8cm,inner sep=0.05cm} }
\tikzset{arc/.style = {->,> = latex, , } }
\begin{figure*}
\centering
    \begin{minipage}{.6\textwidth}
    \scalebox{0.6}{
    \begin{tikzpicture}[auto,node distance = 0.1 cm, scale = 1.4] 
    \node[node] at(-9,-0.75) (S) { New instance };
    \node[node] at(-6.5,-0.75) (R) { Routing Algorithm };
    \node[node] at(-3,0) (H) { Algorithm Decision Maker};
    \node[node] at(-3,-1.) (V) {Human Decision Maker $i$};
    \node[node] at(1., -0.) (Y) { Final Decision };
    \node[node] at(1., -1.) (U) { Unobserved Confounders };
    \draw[arc] (S) to node{ } (R); 
    \draw[arc] (R) to node{ } (H); 
    \draw[arc] (R) to node{ } (V); 
    \draw[arc] (U) to node{ } (V); 
    \draw[arc] (H) to node{ } (Y); 
    \draw[arc] (V) to node{ } (Y); 
    \draw
    (0,-0.75) coordinate (A)
    (3.3,-0.75) coordinate (B)
    (3.3,1) coordinate (C)
    (2.1,1) coordinate (D)
    ;
    \end{tikzpicture}
    } 
    \label{fig:illus2}
    \caption{Human-AI Collaboration  with Unobserved Confounders}
    \end{minipage}
\begin{minipage}{.25\textwidth}
        \includegraphics[width=\textwidth]{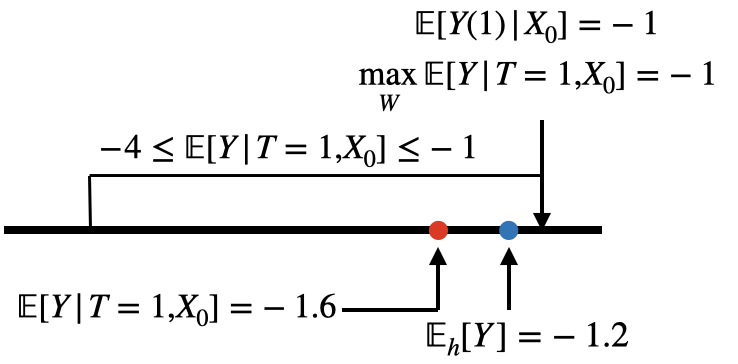}
        \caption{Toy Example}
        \label{fig:toy_illus}
\end{minipage}
\vspace{-1mm}
\end{figure*}

Let $\Tilde{W}_i := \frac{1}{\Tilde{\pi}_0(T_i|X_i)}$ and $W_i := \frac{1}{\pi_0(T_i|X_i,Y_i)}$. By the MSM,  our key observation is that the true weights $W_i$ are bounded in the uncertainty set $\mathcal{W}_n^\Gamma = \{W: 1 + \Gamma^{-1} (\Tilde{W}_i - 1) \leq W_i \leq 1 + \Gamma (\Tilde{W}_i - 1), \forall i=1,\cdots,n\}$. Hence, the worst-case empirical estimator $\hat{R}_n(\pi, \phi, \pi_c, \mathcal{W}_n^\Gamma)=$ 
{\small
\begin{align}
    & \sum_{t=0}^{m-1}\frac{\frac{1}{n}\sum_i  {\mathbb{I}(T_i=t)} [(1-\phi(X_i))\pi(T_i|X_i)-\pi_c(T_i|X_i)] W_i Y_i}{\frac{1}{n}\sum_i \mathbb{I}
    (T_i=t)W_i }  \notag \\
    & + \max_{W}\frac{1}{n}\sum_{i=1}^n \phi(X_i)(Y_i+C(X_i)) \label{eq:emp-r} \\ 
    &\quad\quad\quad ~~~s.t.~~~ 1 + \Gamma^{-1} (\Tilde{W}_i - 1) \leq W_i \leq 1 + \Gamma (\Tilde{W}_i - 1), \notag 
\end{align}
}

The algorithm chooses the policy and router that  minimize the  regret bound $\Bar{\pi}(\Pi, \Phi, \pi_c, \mathcal{W}_n^\Gamma), \Bar{\phi}(\Pi, \Phi, \pi_c, \mathcal{W}_n^\Gamma) = $
\begin{align}
\label{eqn:optimization}
    \underset{\pi\in\Pi, \phi\in\Phi}{\arg\min}\hat{R}_n(\pi, \phi, \pi_c, \mathcal{W}_n^\Gamma).  
\end{align}    

In this case, the algorithm will choose to select the robust routing and decision policy considering that humans may use unobserved information for their decision making. The resulting system is confounding-robust in the sense that it considers the worst risk when humans' behavior cannot be point-identified because of the unobserved confounding. 

Similarly, if we are interested in the policy improvement over the human's policy, we can optimize the future decision and routing policy by minimizing $\hat{R}_n^H(\pi, \phi, \mathcal{W}_n^\Gamma)$ as 
\begin{align}\label{eqn:emp-human}
     \max_{W} &\sum_{t=0}^{m-1}\frac{\frac{1}{n}\sum_i  {\mathbb{I}(T_i=t)} [(1-\phi(X_i))\pi(T_i|X_i)] W_i Y_i}{\frac{1}{n}\sum_i \mathbb{I}
    (T_i=t)W_i }\nonumber \\
    & +\frac{1}{n}\sum_{i=1}^n (\phi(X_i)-1)(Y_i+C(X_i)) .
\end{align}
by removing the baseline policy $\pi_c$ and contrasting the future system's performance with the performance of the human's decision policy with $\phi^H(X)\equiv 1$. 

In practice, we would want the resulting human-AI system to outperform both the incumbent human policy and a candidate algorithmic policy. Then after our system is optimized, we can check whether \Cref{eq:emp-r} and \Cref{eqn:emp-human} are both smaller than 0, which indicates the resulting human-AI system has a better performance compared to human working alone or the candidate algorithm working alone. We  offer a theoretical improvement guarantee in \Cref{sec:improvement}. 

\subsection{An Illustrative Example}
We use a toy example to illustrate how our method works. 
Assume a single context $X=X_0$ and we observe repeated observations for it. $P(U=1|X_0)=P(U=0|X_0)=0.5$. With some abuse of notations, 
$Y(1)=-2, Y(0) =0$ when $U=1$, and $Y(1)=0, Y(0)=-1$ when $U=0$.
Humans follow $P(T=1|X_0,U=1)=0.5+\gamma$ and $P(T=1|X_0,U=0)=0.5-\gamma$ ($\gamma=0$ means no unobserved confounding). $C(x)\equiv 0$. Here the two potential algorithms' performance are %
$\mathbb{E}[Y(1)|X_0]=-1$, $\mathbb{E}[Y(0)|X_0]=-0.5$.

With some simple algebra, the human performance is $\mathbb{E}_h[Y]=-0.75-1.5\gamma$. The nominal propensity score is $P(T=1|X_0)=0.5$. If we want to evaluate the AI policies using observational data (no $U$),  by the inverse propensity score weighting (or from Bayes theorem), 
$\mathbb{E}[Y|T=1,X_0]=$
{\small 
\begin{align}
     \mathbb{E}\frac{\mathbb{I}(T=1,Y=-2)}{P(T=1|X_0)}(-2)
     =\frac{-2P(T=1,U=1 | X_0)}{P(T=1|X_0)}\nonumber 
\end{align}
}
Plugging in the nominal propensity, $\mathbb{E}[Y|T=1,X_0]=-1-2\gamma$. Similarly, $\mathbb{E}[Y|T=0,X_0]=-0.5+\gamma$. When $\gamma=0.3$, $\mathbb{E}[Y|T=1,X_0]=-1.6<-1.2=\mathbb{E}_h[Y]$, so an algorithm assuming unconfoundness will incorrectly think the AI policy $T=1$ is the optimal policy, but human is actually better %
($\mathbb{E}_h[Y]=-1.2<-1=\mathbb{E}[Y(1)|X_0]$). 

Here, the MSM assumption corresponds to $\frac{1}{1+\Gamma}\leq P(T=1|X_0,U)\leq\frac{\Gamma}{1+\Gamma}$, so $\gamma=0.3$ means $\Gamma=4$. Plugging in the confidence interval, we have $-4 \leq \mathbb{E}[Y|T=1,X_0] \leq -\frac{0.5+0.3}{0.8}=-1$.
Since our method adopts the pessimistic principle, the worst risk of the AI algorithm is $-1>\mathbb{E}_h[Y]$, thus our algorithm can choose the best decision maker robustly in the presence of the uncertainty. This is illustrated in \Cref{fig:toy_illus}. ~\looseness=-1

\subsection{Personalization}
In the collaborative objective \Cref{eq:first}, we assume the experts have similar performance. However, this may not be the case in real-world scenarios. Experts often possess different areas of expertise, and may get different levels of confounding information. 
Therefore, implementing a personalized routing model could potentially enhance the performance of the human-machine team. 

Rather than indiscriminately assigning an expert to evaluate a given instance, the routing algorithm can make a decision to either delegate the instance to an algorithm or to a human, and, more importantly, determine the most suitable human decision-maker for the task at hand accounting for varied degrees of confounding for each human. We assume the odds ratio of each human decision-maker $H \in \{0,\cdots,K-1\}$'s propensity scores are associated with the confounding bound $\Gamma_H$. 
Similarly, the policy improvement with personalization has confounding-robust objective ${R}_n^P(\pi, \phi, \pi_c, \mathcal{W}_n^{\Gamma_H})$ is

\begin{align}
    & \mathbb{E}\frac{\phi(H|X)}{d_0(H|X)}(Y+C(X)) \nonumber \\
    & + \sum_{t=0}^{m-1}\frac{\mathbb{E}  \frac{\mathbb{I}(T=t)}{\pi_0(T|X,Y,H)} [\phi(a|X)\pi(T|X)-\pi_c(T|X)] Y }{\mathbb{E}\frac{\mathbb{I}(T=t)}{\pi_0(T|X,Y,H)}} 
\end{align}

Let $\Tilde{W}_i = \frac{1}{\Tilde{\pi}_i(T_i|X_i,H_i)}$, $W_i = \frac{1}{\pi_i(T_i|X_i,Y_i,H_i)}$, then the worst-case estimator $\hat{R}_n^P(\pi, \phi, \pi_c, \mathcal{W}_n^{\Gamma_H})$ is
{\small 
\begin{align}
\label{eqn:jcp_objective}
\scriptstyle
 &\max_{W} \sum_{t=0}^{m-1}\frac{\sum_{i}  \mathbb{I}(T_i=t)W_i [\phi(a|X_i)\pi(T_i|X_i)-\pi_c(T_i|X_i)] Y_i }{\sum_{i}{\mathbb{I}(T_i=t)W_i}} \nonumber \\
 & \qquad + \frac{1}{n}\sum_{i=1}^n\frac{\phi(H_i|X_i)}{d_0(H_i|X_i)}(Y_i+C(X_i)) \\
    &\quad\quad\quad ~~~s.t.~~~ 1 + \Gamma_{H_i}^{-1} (\Tilde{W}_{i} - 1) \leq W_i \leq 1 + \Gamma_{H_i} (\Tilde{W}_i - 1). \nonumber 
\end{align}
}%

The policy and router can be similarly found by optimizing the following objective, $\Bar{\pi}(\Pi, \Phi, \pi_c, \mathcal{W}_n^{\Gamma_{H}}), \Bar{\phi}(\Pi, \Phi, \pi_c, \mathcal{W}_n^{\Gamma_{H}}) =$
\begin{align}
\label{eqn:jcp_optimization}
    \underset{\pi\in\Pi, \phi\in\Phi}{\arg\min}\hat{R}_n^P(\pi, \phi, \pi_c, \mathcal{W}_n^{\Gamma_{H}}). 
\end{align}

Compared to \Cref{eqn:optimization}, \Cref{eqn:jcp_optimization} further considered how to leverage individual human expertise to minimize the human-AI team's risk. When the historical and future human assignment is fully randomized and each human decision maker has the same $\Gamma$, \Cref{eqn:jcp_optimization} recovers \Cref{eqn:optimization}.

\subsection{Implementations}

\textbf{Optimizing the Objectives.~}  
To optimize the deferral collaboration system in \Cref{eqn:optimization} and \Cref{eqn:jcp_optimization}, first, we need to solve the inner maximization in \Cref{eq:emp-r} and \Cref{eqn:jcp_objective}. 

To simplify notations, we consider the following  problem
\begin{align}
\hat{Q_t}(r,\mathcal{W}) = & \max_{W\in\mathcal{W}} \frac{\sum_{i=1}^nr_i W(T_i,X_i,Y_i)}{\sum_{i=1}^n W(T_i,X_i,Y_i)} \notag \\
& s.t. ~~ a_i^{\Gamma_i} \leq W(T_i,X_i,Y_i) \leq b_i^{\Gamma_i}
\label{eq:frac}
\end{align}

When 
$
r_i = \mathbb{I}(T_i=t) [(1-\phi(X_i))\pi(T_i|X_i)-\pi_c(T_i|X_i)] Y_i
$, $a_i^{\Gamma_i} = 1 + \Gamma^{-1} (\Tilde{W}_i - 1)$, $b_i^{\Gamma_i} = 1 + \Gamma (\Tilde{W}_i - 1)$, 
solving \Cref{eq:frac} is equivalent to optimizing $W$ for the empirical $\hat{R}_n(\pi, \phi, \pi_c, \mathcal{W}_n^\Gamma)$ in \Cref{eq:emp-r}, and when 
$r_i = \mathbb{I}(T_i=t) [\phi(a|X_i)\pi(T_i|X_i)-\pi_c(T_i|X_i)] Y_i$, $a_i^{\Gamma_i} = 1 + {\Gamma_{H_i}}^{-1} (\Tilde{W}_i - 1)$, $b_i^{\Gamma_i} = 1 + \Gamma_{H_i} (\Tilde{W}_i - 1)$, solving \Cref{eq:frac} is equivalent to optimizing $W$ for $\hat{R}_n^P(\pi, \phi, \pi_c, \mathcal{W}_n^{\Gamma_H})$ in \Cref{eqn:jcp_objective}. 

The optimization problem in \Cref{eq:frac} is known as a \emph{linear fractional program} \cite{chadha2007linear}. Taking the derivative of the objective in \Cref{eq:frac} w.r.t. $W_i = W(T_i,X_i,Y_i)$, the objective is monotonically increasing (decreasing) with $W_i$ if $r_i\sum_{j \neq i} W_j - \sum_{j \neq i} r_j W_j$ is greater (less) than zero. Hence the optima is achieved when all the $W_i$ are taking the value at the boundary. Furthermore, the objective can be viewed as a weighted combination of $r_i$ with the weights adding up to one. So the objective is maximized when the weights $W_i / \sum_i W_i$  are high for the large $r_i$   and are low for the small $r_i$. Based on these insights, the optimal weights $\{W_i\}$ of the linear fractional program can be characterized by the following theorem. 

\begin{theorem}
\label{thm:frac}
Let $(i)$ be the ordering such that $r_{(1)}\leq r_{(2)} \leq \cdots \leq r_{(n)}$. $\hat{Q_t}(r,\mathcal{W})=\lambda(k^*)$, where $k^*=\inf\{k=1,\cdots, n+1:\lambda(k)<\lambda(k-1)\}$ and 
\begin{align}
    \lambda(k) = \frac{\sum_{i<k}a_{(i)}^\Gamma r_{(i)}+\sum_{i\geq k}b _{(i)}^\Gamma r_{(i)}}{\sum_{i<k}a_{(i)}^\Gamma +\sum_{i\geq k}b _{(i)}^\Gamma }
\end{align}
\end{theorem}

See \Cref{app:proof} for the proof. \Cref{thm:frac} provides an efficient way to solve \Cref{eq:frac} by line search: first sort $r_i$ in ascending order and  initialize all $W_i = a_i^\Gamma$, then change $W_k$ to $b_k^\Gamma$ for $k = n, n-1, \cdots, 1$ until the first time when $\lambda(k)$ decreases. 
After solving the inner maximization problem, we can optimize the minimization problem in \Cref{eqn:optimization} and \Cref{eqn:jcp_optimization}. In this paper, we consider differentiable policies $\Pi = \{\pi_\theta: \theta \in \Theta\}$ and router class $\Phi = \{\phi_\rho, \rho \in \mathrm{P}\}$, 
such as logistic policies with $\pi_{\{\alpha,\beta\}}(x) = \sigma(\alpha + \beta^T x)$ or neural networks, so the following optimization problem can be efficiently solved by gradient descent. For every iteration, our algorithm starts by finding the weights $W$ given the current model parameters through line search, then uses gradient descent to update policy and router jointly. We call our main algorithm assuming all decision makers can only be queried randomly as ConfHAI and its variant considering the diverse expertise of individual human decision makers as ConfHAIPerson.  ~\looseness=-1

In practice, the value of $\Gamma$ can also be estimated in a data-driven way, which we discuss in detail in \Cref{sec:calibration}.

\begin{algorithm}
\caption{Confounding-Robust Deferral Collaboration (ConfHAI/ConfHAIPerson)}\label{alg:hai}
\begin{algorithmic}
\Require  number of iterations $N$, $\pi, \phi$, $\Gamma$ ($\Gamma_H$), $\{X_i,T_i,Y_i\}_{i=1}^N$, $\pi_c$
\Ensure $\pi_\theta, \phi_\rho$
\For{$i \gets 1$ to $N$}
    \State $W \leftarrow \underset{W\in\mathcal{W}}{\arg\max}$ \Cref{eq:emp-r} (\Cref{eqn:jcp_objective} for ConfHAIPerson). 
    \State $\theta, \rho \leftarrow \nabla \hat{R}_n(\pi, \phi, \pi_c, \mathcal{W}_n^\Gamma)$ ($\nabla \hat{R}_n^P(\pi, \phi, \pi_c, \mathcal{W}_n^{\Gamma_H})$ for ConfHAIPerson). 
\EndFor 
\end{algorithmic}
\end{algorithm}

\section{Theoretical Analysis}
\label{sec:improvement}

\noindent \textbf{Improvement Guarantees.~} 
We first show the worst-case empirical regret is an asymptotic upper bound for the population regret. 
We assume the outcome and true propensity score is bounded for analysis, \ie $|Y|\leq B, \pi_0(t|x,y) \geq v, \forall t\in\{0,\cdots,m-1\},x\in\mathcal{X},y\in\mathcal{Y}$. The following theorem guarantees the improvement over the population regret by solving the minimax optimization for the empirical regret. 

\begin{theorem}
\label{thm:improvement}
Suppose the true inverse propensities $1/\pi_0(T_i|X_i,Y_i)\in \mathcal{W}_n^\Gamma, i=1,\cdots, n$, $|Y|\leq B, C(X)\leq \Bar{c}, \pi_0(t|x,y) \geq v, \forall t\in\{0,\cdots,m-1\},x\in\mathcal{X},y\in\mathcal{Y}$ and denote policy and router's class $\Pi$ and $\Phi$'s Rademacher Complexity as $\mathfrak{R}_n(\Pi)$ and $\mathfrak{R}_n(\Phi)$, then for $\delta >0$, 
with probability $1-\delta$, we have 
\begin{align}
    &\resizebox{\hsize}{!}{${R}(\pi, \phi, \pi_c) \leq \hat{R}_n(\pi, \phi, \pi_c, \mathcal{W}_n^\Gamma)  + 2(B+\Bar{c})\mathfrak{R}_n(\Phi).$} \nonumber\\
     & +  
     2\frac{B}{\nu}\mathfrak{R}_n(\Pi) + (3B+3\Bar{c} + \frac{5B+1}{\nu^2})\sqrt{\frac{2\log{\frac{8m}{\delta}}}{n}} \nonumber
\end{align}

\end{theorem}

The proof is included in \Cref{app:proof} and can be  extended for the improvement guarantee over the personalized version of the algorithm. Note that the global optima of the empirical objective is never positive when $\pi_c \in \Pi$ since we can take $\pi = \pi_c$ and $\phi(X)=0$.  If we only consider $\Pi, \Phi$ with vanishing Rademacher Complexity %
(\ie $O(n^{-1/2})$), then \Cref{thm:improvement} implies that given enough samples, if the empirical objective is negative, we can get an improvement over $\pi_c$ under well-specification. We also provide the improvement guarantee over the incumbent human policy in \Cref{cor:humanimp}. 

\begin{corollary}\label{cor:humanimp}
Under the condition of \Cref{thm:improvement} and suppose $\mathfrak{R}_n(\Pi)=O(\frac{1}{\sqrt{n}})$, $\mathfrak{R}_n(\Phi)=O(\frac{1}{\sqrt{n}})$, then for $\delta >0$, 
with probability $1-\delta$, we have 
\begin{align}
    {R}_H(\pi, \phi, \pi_c) \leq \hat{R}_n^H(\pi, \phi, \pi_c, \mathcal{W}_n^\Gamma) +  
     O(\sqrt{\frac{\log{\frac{m}{\delta}}}{n}}).
\end{align}
\end{corollary}

\noindent \textbf{What Kind of Instances are routed to Humans.}
One interesting question under the proposed deferral collaboration framework is 
what kind of instances should be solved by humans and what should be solved by algorithms.
Here we provide a theoretical analysis of the routing decision under the optimal AI policy.
Assume we have access to the true human behavior policy and for an AI policy $\pi(T|X)$ given only $X$, we can compare the expected risk of routing the instance to human and the expected risk of routing the instance to the AI and choose the one with lower expected risk. This produces a closed-form solution for the routing decision. %

\begin{theorem}[Instances routed to the humans]\label{thm:routing}
    Assume we have access to the true weight $W^*$ and an AI policy $\pi(T|X)$ given only $X$, to minimize \Cref{eqn:obj_r}, the routing system should send the decision task to humans when 
    \begin{align}
    \small
        \mathbb{E}_{U\sim P(U|X),T\sim\pi_0(T|X,U)}[Y+C(X)|X] < \mathbb{E}_{T\sim\pi(T|X)}[Y|X] \nonumber 
    \end{align}
\end{theorem}

The proof is in \Cref{app:proof}. The left term is the expected risk of routing the instance to human when humans have access to the unobserved confounder $U$ and the right term is the expected risk of routing the instance to the AI when it only has access to $X$. The theorem has an interesting implication that the routing system should always send the instance to human when humans can utilize $U$ to improve their decision making performance and surpass the best possible decision performance when only $X$ is available. Compared to \citet{gao2021human}, where the main source of complementarity comes from model misspecification, here the main source of complementarity comes from the unobserved confounder and humans may be irreplaceable even if we have access to the optimal AI policy with confounded data.

\section{Experiments}

We report empirical findings to examine the advantages of Human-AI complementary and being robust to unobserved confounding. 
Our first experiment demonstrates the benefit of human-AI collaboration within a controlled environment.
Our subsequent experiments consider two real-world examples in financial lending and healthcare industry.

In our experiment, we examine the following decision-making configurations. For all baselines without using personalization, human experts are selected at random. Human Only (Human) solely queries human decision-makers randomly to output final decisions. Algorithm Only (AO) uses the inverse propensity score weighting method \cite{swaminathan2015counterfactual} to train a policy. 
Confounding-Robust Algorithm Only (ConfAO) trains a confounding-robust policy with no human involved to determine the final decisions \citep{kallus2018confounding}. Human-AI team (HAI) uses the deferral collaboration method assuming unconfoundedness \citep{gao2021human}. %
Our method and its personalized variant are denoted as \emph{ConfHAI} and \emph{ConfHAIPerson} respectively. See \Cref{app:baseline} for a more detailed discussion about the baselines. We use the logistic policies for the policy and router model classes. The baseline policy is set as the never-treat policy $\pi_c(0|x)=1$  \citep{kallus2018confounding}. ~\looseness=-1

\subsection{Synthetic Experiment}
\label{sec:syn_exp}
We demonstrate the benefit of confounding-robust human-AI collaboration using the following data-generating process, %
\begin{align}
    & \resizebox{\hsize}{!}{$\xi \sim \text{Bern}(0.5),  X  \sim \mathcal{N}((2\xi-1)\mu_x, I_5), U = \mathbb{I}[Y_i(1)<Y_i(-1)]$} \nonumber \\
    & \resizebox{\hsize}{!}{$Y(t) = \beta_0^T x + \mathbb{I}[t=1]\beta_{\text{treat}}^T x + 0.5\alpha\xi\mathbb{I}[t=1] + \eta + w\xi + \epsilon$} \nonumber 
\end{align}
where $\beta_{\text{treat}} =  [1.5, 1, 1.5, 1, 0.5]$, $\mu_x = [1, .5, 1, 0, 1]$, $\eta = 2.5, \alpha = -2, w = 1.5$ and $\epsilon\sim\mathcal{N}(0,1)$ \citep{kallus2021minimax}. The nominal propensity $\pi_0(T=1|X) = \sigma(\beta^T X)$, $\beta = [0, .75, .5, 0, 1, 0]$, $T_i$ is generated by the true propensities by $\pi_0(T=1|X,U) = \frac{(\Gamma U + 1-U)\pi_0(T=1|X)}{[1+2(\Gamma-1)\pi_0(T=1|X)-\Gamma]U+\Gamma+(1-\Gamma)\pi_0(T=1|X)}$, where $\Gamma$ is the specified level of confounding. In this setting, the human decision-maker acquires unobserved information to improve decisions. %
We set $\log(\Gamma) = 2.5$, $C(x)=0$ and vary the log-confounding parameter in $\{0.01, 0.5, 1, 1.5, 2, 2.5, 3, 3.5, 4\}$. To also test the personalized variant, we simulate three human decision makers with the same $\Gamma$.  ~\looseness=-1

\begin{figure*}[h]
    \centering
    \begin{subfigure}[b]{0.24\textwidth}
        \includegraphics[width=\textwidth]{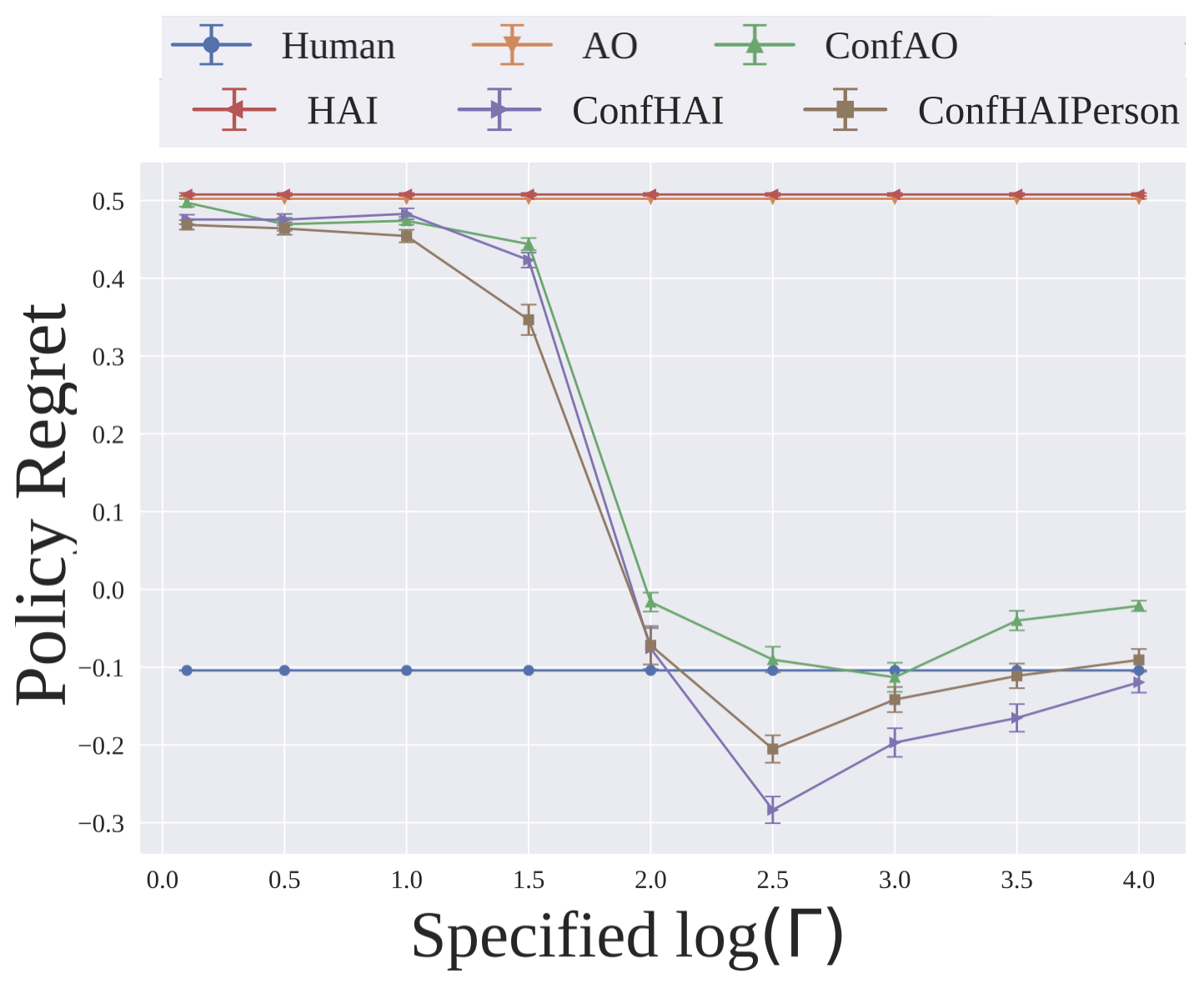}
        \caption{Syn: homogeneous humans.}
        \label{fig:synhomo}
    \end{subfigure}
    \hfill %
    \begin{subfigure}[b]{0.24\textwidth}
        \includegraphics[width=\textwidth]{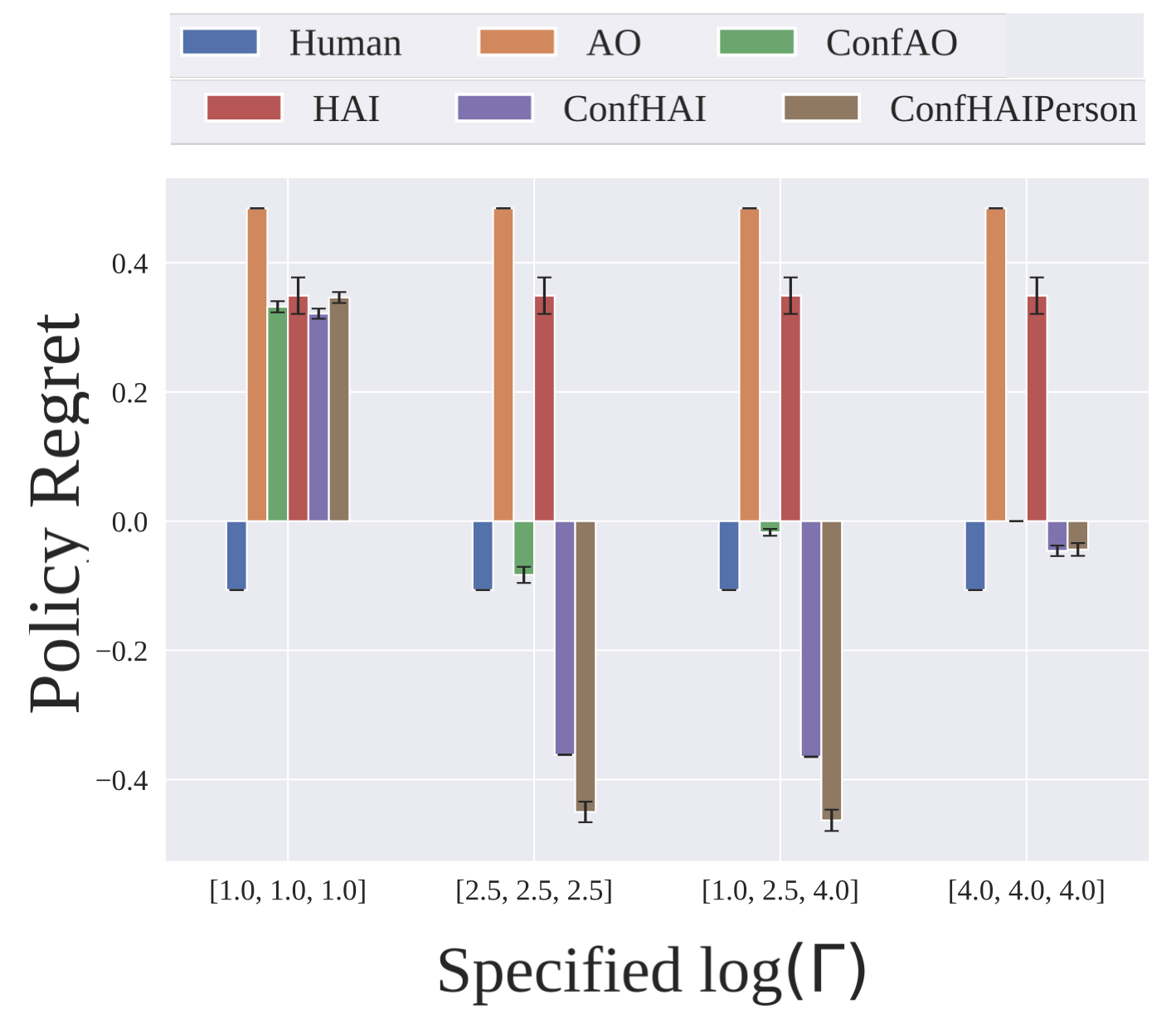}
        \caption{Syn: heterogeous humans.}
        \label{fig:synhetero}
    \end{subfigure}
    \hfill 
    \begin{subfigure}[b]{0.24\textwidth}
\includegraphics[width=\linewidth]{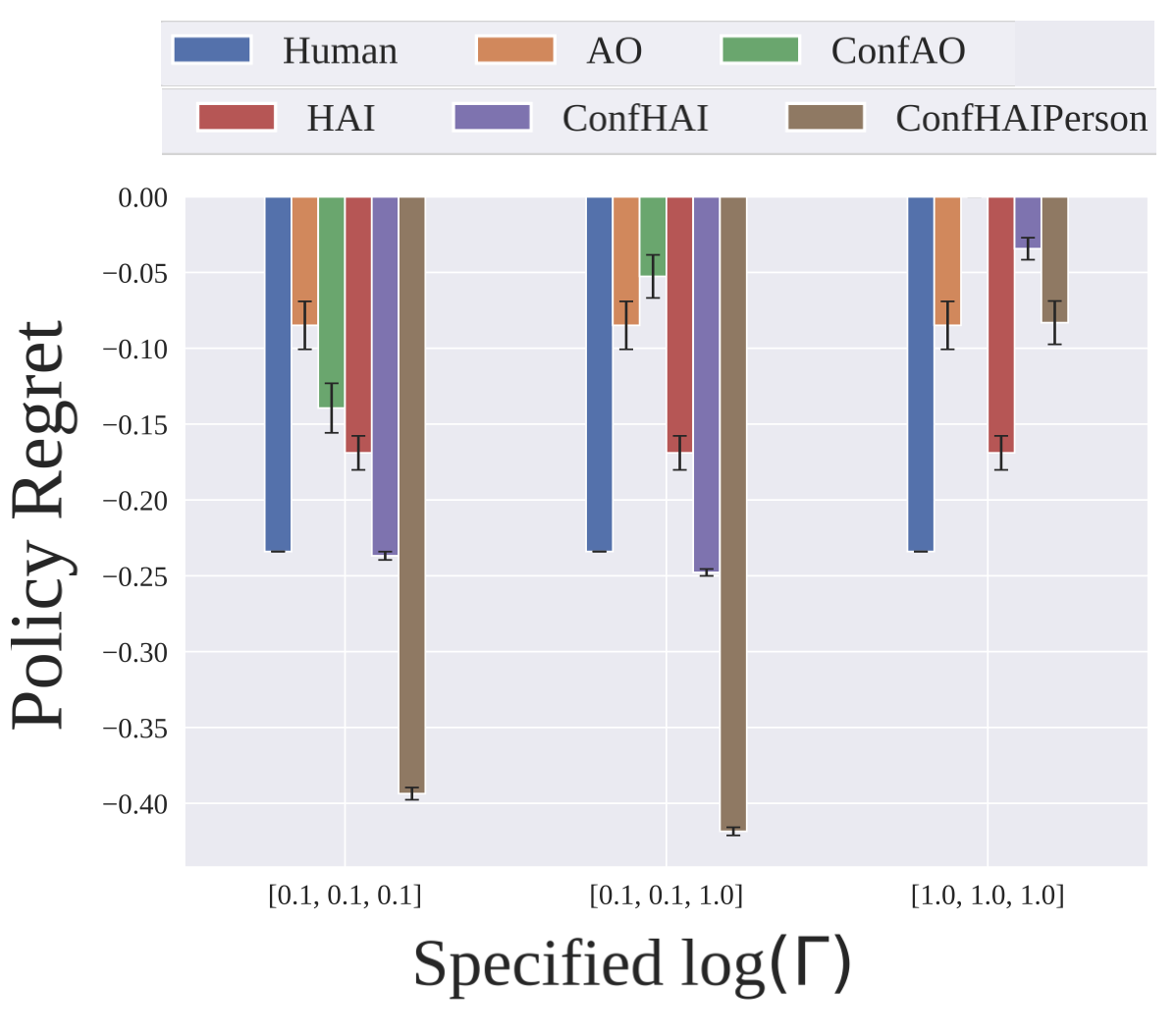}
        \caption{HELOC}
        \label{fig:heloc}
    \end{subfigure}
    \hfill %
    \begin{subfigure}[b]{0.24\textwidth}
        \includegraphics[width=\linewidth]{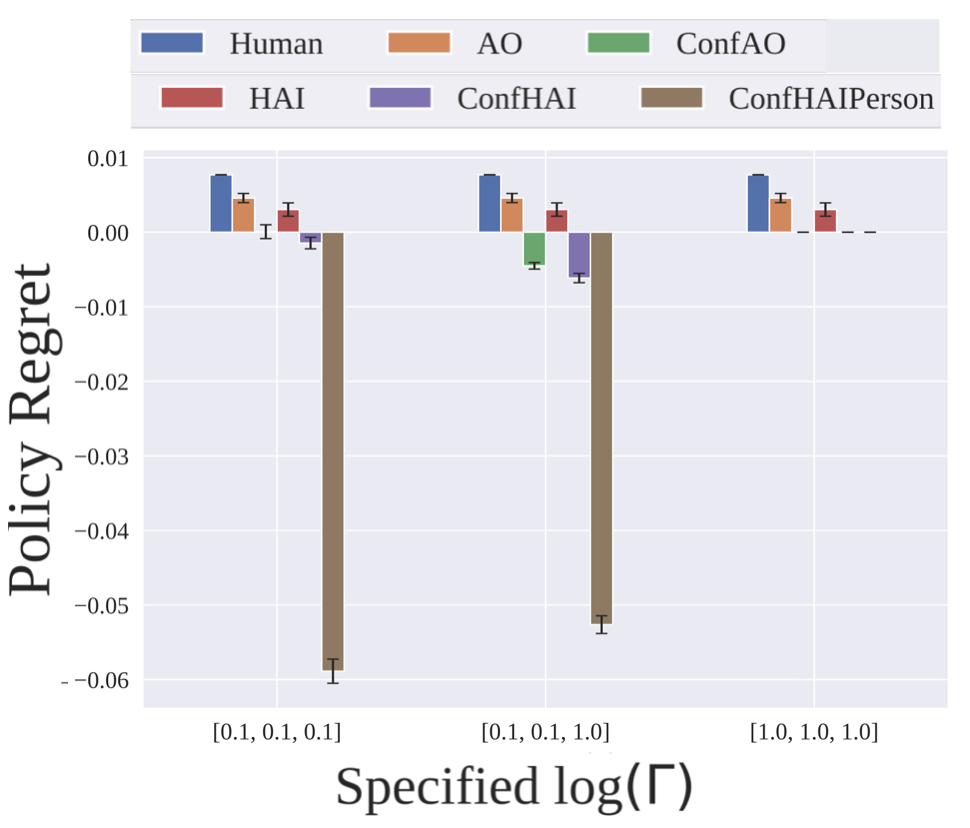}
        \caption{IST}
        \label{fig:ist}
    \end{subfigure}
    \caption{Policy Regret. %
    ConfHAI and ConfHAIPerson offer consistent and significantly better policy improvement for a  range of $\Gamma$
    compared to baseline algorithms over synthetic and real datasets. 
    }
    \label{fig:synexp}
    \vspace{-1mm}
\end{figure*}

The results are shown in \Cref{fig:synhomo}. When $\Gamma$ is small (weak unmeasured confounding), the baselines not considering unobserved confounding have similar performance with our methods and ConfAO. When $\Gamma$ is approaching the underlying confounding factor, we observe a significant policy improvement over the baseline policy (regret is smaller than 0). The personalized method has performance similar to ConfHAI since all human decision makers have the same performance here. 
In this example, interestingly, the ConfAO policy is actually worse than humans' performance, and almost never exceeds humans' performance with varying $\Gamma$, while human-AI complementarity can outperform humans' performance for a range of $\Gamma$ close to its true value, 
which emphasizes the benefit of our confounding-robust deferral system. 

Next, we simulate three human workers with $\log(\Gamma) = \allowbreak 1, 2.5, 4$, respectively, which corresponds to the setting where different humans may acquire different unobserved information to aid their decision making and some experts may perform better than their peers. We examine four $\log(\Gamma)$ specifications with heterogeneous workers: $[1,1,1], [2.5,2.5,2.5], [1,2.5,4], [4,4,4]$ and show the results in \Cref{fig:synhetero}. With small $\Gamma$, we observe all methods perform suboptimally with no policy improvement. 
The personalized variant has an additional improvement over ConfHAI by leveraging the diverse expertise of human decision makers. 
With correctly specified and relatively large $\Gamma$, we observe that ConfHAI and ConfHAIPerson significantly outperform other baselines and demonstrate human-AI complementarity, where they outperform both human-only and algorithm-only teams. ~\looseness=-1

\subsection{Real-World Examples}

We provide two real-world examples in this section with human cost set as $C(x) = 0.1$. See \Cref{app:exp} for more details about the datasets and training in the experiments. %

\noindent \textbf{Financial Lending.} In  financial lending,  loan officers can obtain additional information for decision-making by visiting the loan applicants. However, %
such information may not be recorded in the historical data. 
We use the Home Equity Line of Credit(HELOC) %
dataset which contains anonymized information about credit applications by real homeowners. Some of the used features are the average months since the account opened, maximum delinquency, and number of inquiries in the last 6 months. 
We assume there are three human decision makers with $\log(\Gamma) = [0.1,0.1,1]$, which means two of them rarely seek external information to improve their decision making and another decision maker is more likely to get external risk estimation when evaluating applications. We train a logistic regression on 10\% of the data to simulate nominal policies, which can be a guideline policy of the insurance company, and the actual treatments taken are generated using the same procedure in \Cref{sec:syn_exp} and the fitted nominal propensity is estimated using logistic regression on actual treatments. 
The outcome of the dataset is a binary outcome indicating whether the applicant was 90 days past due. %
We build a risk function where the loan company will receive a risk $Y \sim \mathcal{N}(0,1)$ if not approving the loan, $Y \sim \mathcal{N}(-2,1)$ if approving for an applicant with good credit and $Y \sim \mathcal{N}(2,1)$ if approving for an applicant with bad credit.

\noindent \textbf{Acute Stroke Treatment.}
In this healthcare example, the doctors need to treat patients with acute stroke. Experienced doctors may observe bedside information, and past patient behaviors to aid their decision-making, which are not recorded in the historical records. We use the data from the International Stroke Trial \cite{international1997international} and focus on two treatment arms: the treatment of both aspirin and heparin (medium and high doses), and the treatment of aspirin only. Since the trial only has the outcome under action taken, we create potential outcomes by fitting a separate random forest model for each treatment as in \cite{biggs2021loss,elmachtoub2023balanced}. The outcome is a composite score including variables like death,  recurrent stroke, pulmonary embolism, and recovery. Some of the features used by the algorithm include age, sex, deficient symptoms, stroke types, and cerebellar signs. Similarly, we assume there are three human physicians with $\log(\Gamma) = [0.1,0.1,1]$ prescribing treatments. ~\looseness=-1

The results are shown in \Cref{fig:heloc} and \Cref{fig:ist} respectively. For each experiment, we try three $\log(\Gamma)$ specifications: $[0.1,0.1,0.1]$, $[0.1,0.1,1]$ and $[1,1,1]$, which correspond to under, correct and over specifications. In HELOC, the baselines not considering unobserved confounding can still achieve policy improvement, while it is not consistent across settings, \eg in IST. We observe that ConfHAI and ConfHAIPerson achieve the best performance with correctly-specified $\Gamma$ and the personalized variant achieves significantly better performance than other methods. With over-specification, the performance of confounding-robust methods decreases but can reliably provide policy improvement. Similarly, ConfAO can provide policy improvement with different specifications of $\Gamma$, however, its performance is often much worse than the human-AI methods we propose.

\subsection{Real Human Responses}
In addition, we use the real human responses to validate our approach. 
We use the scientific annotation dataset FOCUS \cite{rzhetsky2009get} with responses from five human annotators. The features are sentences extracted from a scientific corpus, and each labeler was asked to label the sentence as scientific or not. We transform the text into feature representations using Glove embeddings \cite{pennington2014glove}. We assume if the human annotator considers the sentence scientific, they will apply action I (e.g., retweet the paper), if the sentence is indeed scientific, the risk is $\mathcal{N}(-1,1)$, otherwise the risk is from $\mathcal{N}(1,1)$. On the other hand, if the human annotator considers the sentence as non-scientific, they will apply action II (e.g., ignore the paper), if the sentence is indeed non-scientific, the risk is  $\mathcal{N}(-1,1)$ (otherwise the risk is $\mathcal{N}(1,1)$). Since each sentence is annotated, we know whether the human annotator considers the sentence scientific and is able to derive the simulated human behavior policy. The confounding is created by removing samples with 20\% top outcomes in the treated group and 20\% bottom outcomes in the control group. %
We specify the same $\Gamma$ for each human and vary it.  

The results are shown in \Cref{fig:real_ablation}. This dataset is different from our simulations since humans' true propensities may not reflect the worst case indicated in the MSM optimization. 
However, we still observe our methods consistently offer the best performance with a wide range of $\Gamma$.

\subsection{Ablation Studies}
\begin{figure}[h]
    \centering
    \begin{subfigure}{0.22\textwidth}
\includegraphics[width=\linewidth]{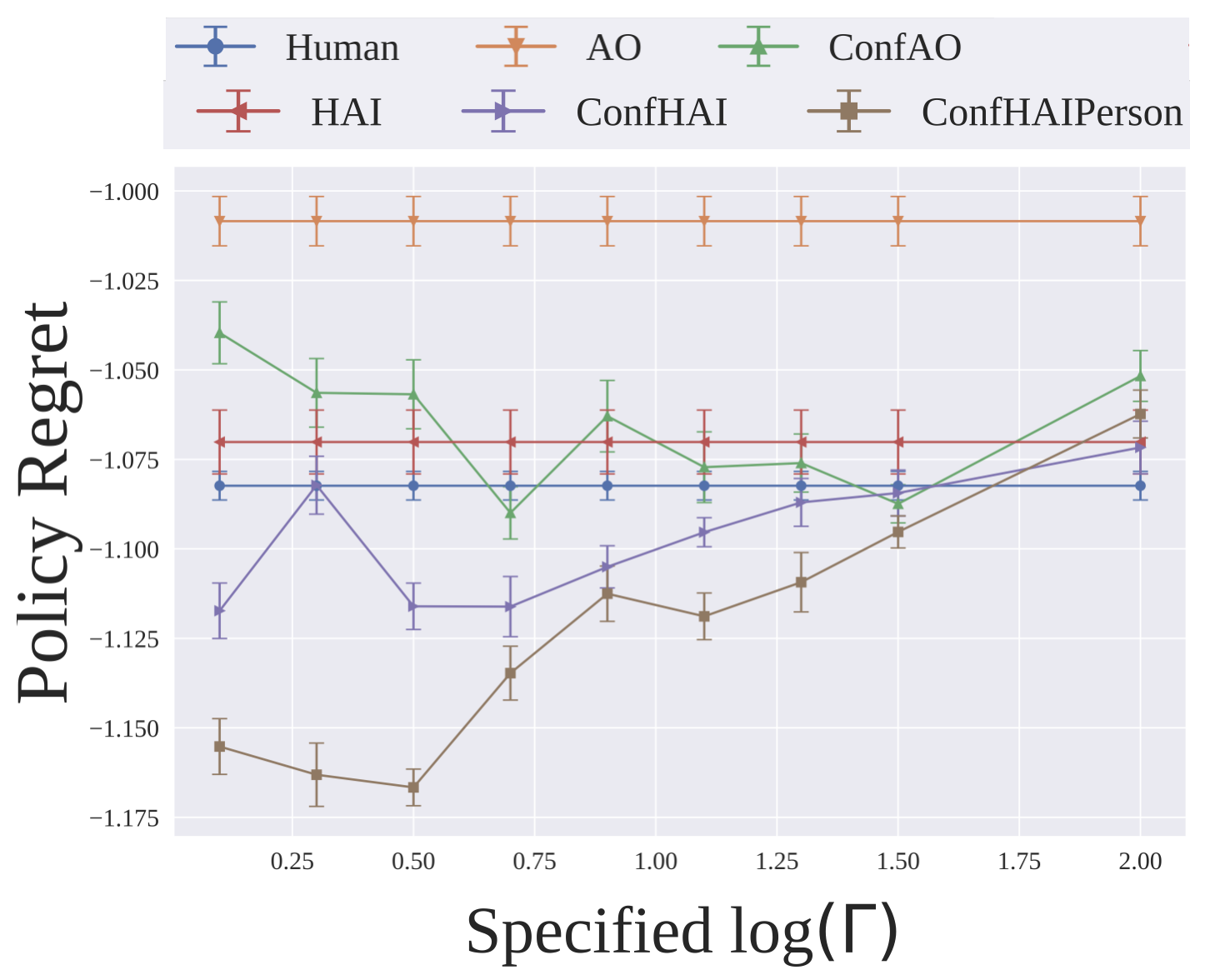}
    \end{subfigure}
    \begin{subfigure}{0.22\textwidth}
        \includegraphics[width=\linewidth]{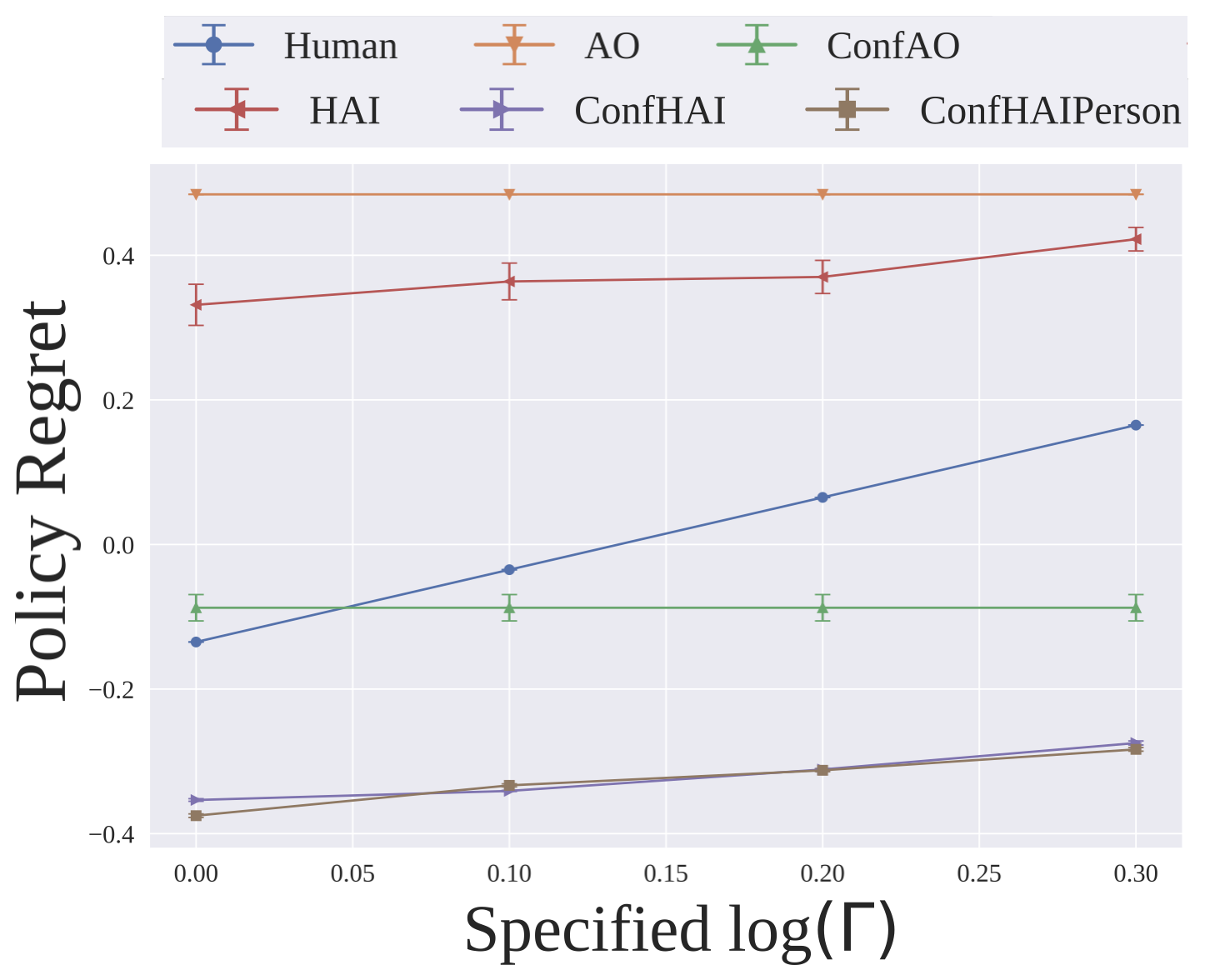}
    \end{subfigure}
    \caption{Real Data (left) and Ablation Studies (right).}
    \label{fig:real_ablation}
      \vspace{-3mm}
\end{figure}

We examine the effect of human cost on the risk of each method in \Cref{fig:real_ablation}. We use the synthetic data setup and vary the human cost from 0 to 0.3. As the cost becomes higher, the Human baseline's performance gets worse. Human-AI systems' performance (HAI, ConfHAI, ConfHAIPerson) is also impacted when human cost is higher, but the proposed methods consistently outperform other baselines.
\section{Conclusion and Future Work}

In this paper, we study a new problem of unmeasured confounding in Human-AI collaboration. We propose ConfHAI as a novel confounding-robust deferral policy learning method to address this problem. ConfHAI optimizes policy decisions by selectively deferring decision instances to either humans or the AI, based on the context and capabilities of both and the strength of the unmeasured confounding. We demonstrate the policy improvements of ConfHAI in theory and through a variety of synthetic and real data simulations. Nevertheless, a potential limitation of the proposed method is the relatively strict constraints on the marginal sensitivity model. Future work can improve the sharpness of the MSM \citep{dorn2023sharp} or consider adopting more interpretable sensitivity models \citep{imbens2003sensitivity}. Broadly speaking, our findings indicate the importance of explicitly accounting for the information discrepancy between human decision-makers and AI algorithms to improve human-AI complementarity.

\bibliography{aaai25}

\newpage 
\onecolumn 
\appendix
\setcounter{secnumdepth}{1}

\begin{center}
	{\huge \textbf{Supplementary Materials}}
\end{center}
\vspace{5mm}
\section{Proofs}
\label{app:proof}

Proof of Theorem 1: 
\begin{proof}
Assume $|Y|\leq B, W\leq 1/\nu$.
Let $\hat{D}_t = \mathbb{E}_n [W^* \mathbb{1}(T=t)]$. With some abuse of notations, we use $\hat{R}(\pi, \phi, \pi_c, \mathcal{W}^*)$ to represent the empirical estimator with the optimal weight $W^*$. 

\begin{align}
    &\sup_{\pi\in\Pi,\phi\in\Phi} {R}(\pi, \phi, \pi_c)-\hat{R}(\pi, \phi, \pi_c, \mathcal{W}^*) \\
    =& \sup_{\pi\in\Pi,\phi\in\Phi} -\frac{1}{n}\sum_i \phi(X_i)(Y_i+C(X_i)) + \mathbb{E}\phi(X)(Y+C(X))\\
    & - \frac{1}{n}\frac{\sum_i \big(\pi(T_i|X_i) (1-\phi(X_i))-\pi_c(T_i|X_i,U_i)\big) W_i Y_i}{\mathbb{E}\hat{D}_{T_i}}(\frac{\mathbb{E}\hat{D}_{T_i}-\hat{D}_{T_i}}{\hat{D}_{T_i}}+1) \\ 
    & + \mathbb{E} \big(\pi(T|X) (1-\phi(X))-\pi_c(T|X,U)\big) W Y \\
    \leq & \sup_{\phi\in\Phi} -\frac{1}{n}\sum_i \phi(X_i)(Y_i+C(X_i)) + \mathbb{E}\phi(X)(Y+C(X))\\
    &+ \sup_{\pi\in\Pi,\phi\in\Phi} (\mathbb{E}_n-\mathbb{E}) \big(\pi(T|X) (1-\phi(X))-\pi_c(T|X,U)\big) W Y + \frac{B}{\nu} \sum_i \frac{1}{n} \frac{\mathbb{E}\hat{D}_{T_i}-\hat{D}_{T_i}}{\hat{D}_{T_i}}
\end{align}

Write 
\begin{align}
     D = \sup_{\phi\in\Phi} \{-\frac{1}{n}\sum_i \phi(X_i)(Y_i+C(X_i)) + \mathbb{E}\phi(X)(Y+C(X)) \}, 
\end{align}
Since it has bounded difference of $\frac{B+\Bar{c}}{n}$, McDiarmid' inequality indicates with probability $1-p_1$, 
\begin{align}
    D-\mathbb{E}D \leq (B+\Bar{c})\sqrt{\frac{\log{\frac{1}{p_1}}}{2n}}
\end{align}
By the standard symmetrization method, we have  

\begin{align}
    \mathbb{E} D \leq 2 \mathbb{E}\sup_{\phi\in\Phi} \sum_i^n \epsilon_i [\phi(X_i)(Y_i  + C(X_i))]
\end{align}

By Rademacher comparison theorem / contraction principle \cite{ledoux1991probability},  

\begin{align}
    \mathbb{E} D \leq  2(B+\Bar{c})\mathbb{E}\mathfrak{R}_n(\Phi)
\end{align}

Next we bound $\frac{B}{\nu} \sum_i \frac{1}{n} \frac{\mathbb{E}\hat{D}_{T_i}-\hat{D}_{T_i}}{\hat{D}_{T_i}}$. Let $n_t=\sum_i \mathbb{1}[T_i=t]$, 
\begin{align}
    \frac{B}{\nu} \sum_i \frac{1}{n} \frac{\mathbb{E}\hat{D}_{T_i}-\hat{D}_{T_i}}{\hat{D}_{T_i}} \leq \frac{B}{n\nu} \sum_t \frac{|\hat{D}_{t}-1|}{\hat{D}_{t}}
\end{align}

For any $t$, write $\rho_t=P(T=t)$, by Hoeffding inequality, 

\begin{align}
    P(|\frac{n_t}{n}-\rho_t|\geq \rho_t/2) \leq 2 \exp(-\nu^2\rho_t^2 n/2)
\end{align}

Take a union bound over $m$ treatments, with probability $p_2$, such that $\frac{1}{\nu}\sqrt{\frac{\log(2m/p_2)}{2n}}\leq \rho_t^2/2$. 

Again with Hoeffding inequality, 

\begin{align}
    P(|\hat{D}_{t}-1|\geq \epsilon)\leq 2\exp(-2\nu^2\epsilon^2n)
\end{align}

With $p_3$, such that $\frac{1}{\nu}\sqrt{\frac{\log(2m/p_3)}{2n}}\leq 1$, thus with $1-p_3$, $\frac{1}{\hat{D}_{t}}\leq 2$ and $\frac{|\hat{D}_{t}-1|}{\hat{D}_{t}} \leq \frac{2}{\nu}\sqrt{\frac{\log(2m/p_3)}{2n}}$

write 
\begin{align}
V = \sup_{\pi\in\Pi,\phi\in\Phi}(\mathbb{E}_n- \mathbb{E}) \big(\pi(T|X) (1-\phi(X))-\pi_c(T|X)\big) W Y    
\end{align}

Since it has bounded difference of $\frac{B}{n\nu}$, McDiarmid' inequality indicates with probability $1-p_4$, 
\begin{align}
    V-\mathbb{E}V \leq \frac{B}{\nu}\sqrt{\frac{\log{\frac{1}{p_4}}}{2n}}
\end{align}
By the symmetrization argument, 

\begin{align}
    \mathbb{E} V \leq 2 \mathbb{E} \sup_{\pi\in\Pi,\phi\in\Phi} \frac{1}{n}\sum_i \epsilon_i  \big(\pi(T_i|X_i) (1-\phi(X_i))-\pi_c(T_i|X_i)\big) W_i Y_i
\end{align}

By Rademacher comparison theorem / contraction principle (\cite{ledoux1991probability}, Thm 4.12), 

\begin{align}
    \mathbb{E} V \leq  2\frac{B}{\nu}\mathbb{E}\mathfrak{R}_n(\Pi)
\end{align}

Finally, with probability $1-p_5$, since $\mathfrak{R}_n(\Pi)$ has bounded differences with $\frac{2}{n}$, by McDiarmid’s inequality we have 

\begin{align}
    \mathbb{E}\mathfrak{R}_n(\Pi) - \mathfrak{R}_n(\Pi) \leq \sqrt{\frac{2}{n}\log{\frac{1}{p_5}}}.
\end{align}

Similarly, with probability $1-p_6$, we have 
\begin{align}
    \mathbb{E}\mathfrak{R}_n(\Phi) - \mathfrak{R}_n(\Phi) \leq \sqrt{\frac{2}{n}\log{\frac{1}{p_6}}}.
\end{align}

Let $p_1=p_2=p_3=p_4=p_5=p_6=\frac{\delta}{6}$, 
\begin{align}
    &\sup_{\pi\in\Pi,\phi\in\Phi} {R}(\pi, \phi, \pi_c)-\hat{R}(\pi, \phi, \pi_c, \mathcal{W}^*)\\ 
    &\leq (3B+3\Bar{c} +\frac{1}{\nu}+\frac{2B}{\nu^2} + \frac{3B}{\nu})\sqrt{\frac{2\log{\frac{\max(6,8m)}{\delta}}}{n}}+ 2\frac{B}{\nu}\mathfrak{R}_n(\Pi) + 2(B+\Bar{c})\mathfrak{R}_n(\Phi) \\ 
    & \leq (3B+3\Bar{c} +\frac{5B+1}{\nu^2})\sqrt{\frac{2\log{\frac{8m}{\delta}}}{n}}+ 2\frac{B}{\nu}\mathfrak{R}_n(\Pi) + 2(B+\Bar{c})\mathfrak{R}_n(\Phi) 
\end{align}

Since we have the well-specificaition assumpition, thus $\hat{R}(\pi, \phi, \pi_c, \mathcal{W}^\star) \leq \hat{R}(\pi, \phi, \pi_c, \mathcal{W}_n^\Gamma)$, which completes the proof. 
\end{proof}

Proof of Corollary 2:
This is a direct result by applying the same arguments in Theorem 1.

Proof of Theorem 3.
\begin{proof}
We prove the claim by contradiction. Suppose $r_1 \leq \cdots \leq r_n$, and suppose the optima $W^*$ has $W^*_i = b_i$, $W^*_j =a_j$ with $i<j$. It means
\begin{align*}
0 < & \frac{\partial}{\partial W_i}\frac{\sum_{i=1}^nr_i W_i}{\sum_{i=1}^n W_i}  \\
= & \sum_{k \neq i,j} (r_i - r_k) W_k + (r_i - r_j) W_j \\
\leq & \sum_{k \neq i,j} (r_j - r_k) W_k + (r_j - r_i)W_i \\
= & \frac{\partial}{\partial W_j}\frac{\sum_{i=1}^nr_i W_i}{\sum_{i=1}^n W_i},
\end{align*}
where the inequality is because $r_i \leq r_j$, $(r_i - r_j)W_j \leq 0 \leq (r_j - r_i)W_i$. The partial derivative means the objective increases with $W_j$, which contradicts the assumption that the optimal $W_j^*$ equals its minimal value $a_j$.  Therefore, for the optimal $W^*$, there exists $k\in \{1,\cdots,n\}$ such that $W^*_i = b_i$ for $i \geq k$ and $W^*_i = a_i$ for $i < k$. 
\end{proof}

Proof of Theorem 4:
\begin{proof}
Since we have access to the true weight $W^*$, we can write \Cref{eqn:obj_r} as 
\begin{align}
    \mathbb{E}_{U\sim P(U|X), T\sim \pi_0(T|X,U)}\phi(X)(Y+C(X)) + \mathbb{E}_{T\sim \pi(T|X)} Y [(1-\phi(X))] - \mathbb{E}_{T\sim \pi_c(T|X)} Y 
\end{align}

It is easy to see it is minimized when we set $\phi(x)=1$ if $\mathbb{E}_{U\sim P(U|X), T\sim \pi_0(T|X,U)}(Y+C(X)) < \mathbb{E}_{T\sim \pi(T|X)} Y $ and $\phi(x)=0$ otherwise.
\end{proof}

\section{Data-Driven Calibration of $\Gamma$.}
\label{sec:calibration}
Since the theorem assumes well-specification, the practitioner needs to specify a plausible strength of the unmeasured confounding by the parameter $\Gamma$. Though the exact value of $\Gamma$ involves the nonidentifiable true propensity score, we can have a reference point by quantifying the impact of the observed covariates on the propensity score. The parameter $\Gamma$ defined in \Cref{eqn:sensitivity} measures the degree of influence of the confounder $U=Y(t)$ on the propensity odds. %
Similar to \citet{hsu2013calibrating}, we measure the impact of an observed covariate $Z$ given the other observed covariates $X \backslash Z$ as a reference. The reference points of $\Gamma$ is computed from the odds ratio $\frac{(1-\tilde{\pi}_0(T|X \backslash Z)\tilde{\pi}_0(T|X)}{\tilde{\pi}_0(T|X \backslash Z)(1-\tilde{\pi}_0(T|X))}$. The covariate $Z$ can also be a group of correlated observed variables that better mimic the nature of unmeasured confounding variables \citep{veitch2020sense}. We refer to \citet{cinelli2020making} for a detailed discussion on potential conservativeness in this calibration procedure. 

\section{Baselines}
\label{app:baseline}
Here we include more details about the baselines we used in the experiments. 

\noindent\textbf{Human Only (Human)}: Human only solely queries human-decision-makers randomly to output final decisions and each human decision maker will incur a cost of $C(X_i)$. 

\noindent\textbf{Algorithm Only (AO)}: uses the inverse propensity score weighting method to train a policy. AO solves the following optimization problem:

\begin{align}
    \min_{\pi\in\Pi} \frac{1}{N}\sum_{i=1}^N\frac{\pi(T_i|X_i)}{\hat{\pi}_0(T_i|X_i)}Y_i, 
\end{align}
which can be efficiently solved by gradient descent for differtiable policy classes. 

\noindent\textbf{Confounding-Robust Algorithm Only (ConfAO)} trains a confounding-robust policy \citep{kallus2018confounding} to determine the final decisions. ConfAO solves a constraint optimization of the AO baseline with the marginal sensitivity model constraint on the nominal propensity scores. More specifically, it optimizes the following objective:

\begin{align}
    &\quad \quad \min_{\pi\in\Pi} \max_{\pi_0} \frac{1}{N}\sum_{i=1}^N\frac{\pi(T_i|X_i)}{{\pi}_0(T_i|X_i,Y_i)}Y_i    \\
    &s.t. \quad \Gamma_i^{-1} \leq  \frac{(1-\tilde{\pi}_0(T_i|X_i)\pi_0(T_i|X_i,Y_i)}{\tilde{\pi}_0(T_i|X_i)(1-\pi_0(T_i|X_i,Y_i))} \leq \Gamma_i.
\end{align}

Similar to our method, it also requires a pre-specified $\Gamma$ to ensure robustness against unobserved confounding, while not considering the possibility for human to make future decisions.  

\noindent\textbf{Human-AI team (HAI)} uses the deferral collaboration method proposed in \cite{gao2021human} to train a router and policy jointly assuming unconfoundness. It optimizes the policy and router jointly using the objective:

\begin{align}
\min_{\phi\in\Phi,\pi\in\Pi}\sum_{i=1}^N \phi(X_i)(Y_i + C(X_i))  + \frac{  (1-\phi(X_i))\pi(T_i|X_i)}{\hat{\pi}_{0}(T_i|X_i)}Y_i. 
\end{align}

\section{Additional Details about Experiments}
\label{app:exp}

In this section, we provide more details about the datasets we used. We use the logistic policies for the policy and router model classes.

\noindent \textbf{Financial Lending.} For HELOC dataset, we use the following features in the experiments: number of months that have elapsed since first trade, number of months that have elapsed since last opened trade, average months in file, number of satisfactory trades, number of trades which are more than 60 past due, number of trades which are more than 90 past due, percent of trades, that were not delinquent, number of months that have elapsed since last delinquent trade, the longest delinquency period in last 12 months, the longest delinquency period, total number of trades, number of trades opened in last 12 months, percent of installments trades, months since last inquiry (excluding last 7 days), number of inquiries in last 6 months, number of inquiries in last 6 months (excluding last 7 days), revolving balance divided by credit limit, installment balance divided by original loan amount, number of revolving trades with balance, number of installment trades with balance, number of trades with high utilization ratio (credit utilization ratio - the amount of a credit card balance compared to the credit limit), and the percent of trades with balance.
The outcome of the dataset is a binary outcome indicating whether the applicant was 90 days past due since the account was opened over 24 months. HELOC dataset has 10459 observations, 5000 of which has `good`' credit and ``5459''has bad credit. We use 10\% of the data to generate the observational data and the rest as test data with 10 random partitions. 

\noindent \textbf{Acute Stroke Treatment.}
For the International Stroke Trial data, we use the following features: 
age, sex, conscious state, systolic blood pressure, pulmonary embolism, deep vein thrombosis, infarct visible on CT, face deficit, arm/hand deficit, leg/foot deficit, dysphasia hemianopia, visuospatial disorder, brainstem/cerebellar signs, other deficit and stroke type. 
Our scalarized composite score is:
\begin{align}
    Y = &2\mathbb{I}[\text{death at discharge}] + \mathbb{I}[\text{recurrent stroke}] + 0.5\mathbb{I}[\text{pulmonary embolism or intracranial bleeding}] \nonumber \\ 
& + 0.5 \mathbb{I}[\text{other side effects}]-2\mathbb{I}[\text{full recovery}]-\mathbb{I}[\text{discharge}] \nonumber 
\end{align}
For categorical variables, we use one-hot encodings as features, which results in 42 features in total. The dataset has 19435 patients with acute ischaemic stroke, since we only focus on patients who are assigned aspirin with median, high, or none heparin dose, it results in 2430 samples in the treated group (more aggressive treatment) and 4858 samples in the control treatment (only aspirin). We randomly split data into 50\%/50\% train-test split with 10 runs.

\section{Notations}
\begin{table}[ht]
\centering
\begin{tabular}{p{8cm}p{6cm}}
\toprule
$\pi_0(t|x,y) = P(T=t|X=x,Y(t)=y)$  and \\  $\pi_0(t|x, y, h) = P(T=t|X=x,Y(t)=y,  H=h)$     & Behavior policy, true propensity \\
$\phi(X): \mathcal{X}\rightarrow [0,1]$ and $d_0(X):  \mathcal{X} \to \Delta^K$& Routing policy \\
$\pi_c(T|X)$ & Baseline policy \\
$\tilde{\pi}_0(t|x) = P(T=t|X=x)$  & Nominal Propensity score \\
$\Tilde{W}_i = \frac{1}{\Tilde{\pi}_0(T_i|X_i)}$ , $W_i = \frac{1}{\pi_0(T_i|X_i,Y_i)}$ \\ 
$W = \{W_1, W_2, \cdots, W_n\}$ & Weights \\ 
$C(X)$ & Human decision-maker cost \\ 
$X_i\in\mathcal{X}$ & Observed covariates \\
$T_i \in \{0,\cdots, m-1\}$ & Treatment \\
$Y_i\in\mathbb{R}$ & Risk \\ 
$U_i \in\mathcal{U}$  & Unobserved confounders \\
\bottomrule
\end{tabular}
\caption{Summary of Notations \label{tab:def}}
\end{table}

\end{document}